\documentclass[aps, twocolumn, superscriptaddress, showpacs, nofootinbib]{revtex4-1}

\usepackage{lipsum}
\usepackage[utf8]{inputenc}
\usepackage[T1]{fontenc}
\usepackage{ae,aecompl} 
\usepackage{graphicx}
\usepackage{amsmath}
\usepackage{color}
\usepackage{amssymb}
\usepackage{latexsym}
\usepackage{wasysym}
\usepackage{psfrag}
\usepackage{comment}
\usepackage{ifthen}
\usepackage[citecolor=blue,colorlinks=true]{hyperref}
\usepackage{longtable}
\usepackage{float}
\usepackage[utf8]{inputenc}
\usepackage{lineno}
\usepackage{subfigure}

\begin{document}

\title{All-sky search for short gravitational-wave bursts in the first Advanced LIGO run}




\author{%
B.~P.~Abbott,$^{1}$  
R.~Abbott,$^{1}$  
T.~D.~Abbott,$^{2}$  
M.~R.~Abernathy,$^{3}$  
F.~Acernese,$^{4,5}$ 
K.~Ackley,$^{6}$  
C.~Adams,$^{7}$  
T.~Adams,$^{8}$ 
P.~Addesso,$^{9}$  
R.~X.~Adhikari,$^{1}$
% ...
\\
\medskip
(LIGO Scientific Collaboration and Virgo Collaboration) 
\\
\medskip
{{}$^{*}$Deceased, March 2016. }%
}\noaffiliation
\affiliation {LIGO, California Institute of Technology, Pasadena, CA 91125, USA }
\affiliation {Louisiana State University, Baton Rouge, LA 70803, USA }
\affiliation {American University, Washington, D.C. 20016, USA }
\affiliation {Universit\`a di Salerno, Fisciano, I-84084 Salerno, Italy }
\affiliation {INFN, Sezione di Napoli, Complesso Universitario di Monte S.Angelo, I-80126 Napoli, Italy }
\affiliation {University of Florida, Gainesville, FL 32611, USA }
\affiliation {LIGO Livingston Observatory, Livingston, LA 70754, USA }
\affiliation {Laboratoire d'Annecy-le-Vieux de Physique des Particules (LAPP), Universit\'e Savoie Mont Blanc, CNRS/IN2P3, F-74941 Annecy-le-Vieux, France }
\affiliation {University of Sannio at Benevento, I-82100 Benevento, Italy and INFN, Sezione di Napoli, I-80100 Napoli, Italy }
\affiliation {Albert-Einstein-Institut, Max-Planck-Institut f\"ur Gravi\-ta\-tions\-physik, D-30167 Hannover, Germany }
\affiliation {Nikhef, Science Park, 1098 XG Amsterdam, The Netherlands }
\affiliation {LIGO, Massachusetts Institute of Technology, Cambridge, MA 02139, USA }
\affiliation {Instituto Nacional de Pesquisas Espaciais, 12227-010 S\~{a}o Jos\'{e} dos Campos, S\~{a}o Paulo, Brazil }
\affiliation {INFN, Gran Sasso Science Institute, I-67100 L'Aquila, Italy }
\affiliation {INFN, Sezione di Roma Tor Vergata, I-00133 Roma, Italy }
\affiliation {Inter-University Centre for Astronomy and Astrophysics, Pune 411007, India }
\affiliation {University of Wisconsin-Milwaukee, Milwaukee, WI 53201, USA }
\affiliation {Leibniz Universit\"at Hannover, D-30167 Hannover, Germany }
\affiliation {Universit\`a di Pisa, I-56127 Pisa, Italy }
\affiliation {INFN, Sezione di Pisa, I-56127 Pisa, Italy }
\affiliation {Australian National University, Canberra, Australian Capital Territory 0200, Australia }
\affiliation {California State University Fullerton, Fullerton, CA 92831, USA }
\affiliation {LAL, Univ. Paris-Sud, CNRS/IN2P3, Universit\'e Paris-Saclay, F-91898 Orsay, France }
\affiliation {Chennai Mathematical Institute, Chennai 603103, India }
\affiliation {Universit\`a di Roma Tor Vergata, I-00133 Roma, Italy }
\affiliation {Universit\"at Hamburg, D-22761 Hamburg, Germany }
\affiliation {INFN, Sezione di Roma, I-00185 Roma, Italy }
\affiliation {Albert-Einstein-Institut, Max-Planck-Institut f\"ur Gravitations\-physik, D-14476 Potsdam-Golm, Germany }
\affiliation {APC, AstroParticule et Cosmologie, Universit\'e Paris Diderot, CNRS/IN2P3, CEA/Irfu, Observatoire de Paris, Sorbonne Paris Cit\'e, F-75205 Paris Cedex 13, France }
\affiliation {West Virginia University, Morgantown, WV 26506, USA }
\affiliation {Universit\`a di Perugia, I-06123 Perugia, Italy }
\affiliation {INFN, Sezione di Perugia, I-06123 Perugia, Italy }
\affiliation {European Gravitational Observatory (EGO), I-56021 Cascina, Pisa, Italy }
\affiliation {Syracuse University, Syracuse, NY 13244, USA }
\affiliation {SUPA, University of Glasgow, Glasgow G12 8QQ, United Kingdom }
\affiliation {LIGO Hanford Observatory, Richland, WA 99352, USA }
\affiliation {Wigner RCP, RMKI, H-1121 Budapest, Konkoly Thege Mikl\'os \'ut 29-33, Hungary }
\affiliation {Columbia University, New York, NY 10027, USA }
\affiliation {Stanford University, Stanford, CA 94305, USA }
\affiliation {Universit\`a di Padova, Dipartimento di Fisica e Astronomia, I-35131 Padova, Italy }
\affiliation {INFN, Sezione di Padova, I-35131 Padova, Italy }
\affiliation {Nicolaus Copernicus Astronomical Center, Polish Academy of Sciences, 00-716, Warsaw, Poland }
\affiliation {Center for Relativistic Astrophysics and School of Physics, Georgia Institute of Technology, Atlanta, GA 30332, USA }
\affiliation {University of Birmingham, Birmingham B15 2TT, United Kingdom }
\affiliation {Universit\`a degli Studi di Genova, I-16146 Genova, Italy }
\affiliation {INFN, Sezione di Genova, I-16146 Genova, Italy }
\affiliation {RRCAT, Indore MP 452013, India }
\affiliation {Faculty of Physics, Lomonosov Moscow State University, Moscow 119991, Russia }
\affiliation {SUPA, University of the West of Scotland, Paisley PA1 2BE, United Kingdom }
\affiliation {Caltech CaRT, Pasadena, CA 91125, USA }
\affiliation {University of Western Australia, Crawley, Western Australia 6009, Australia }
\affiliation {Department of Astrophysics/IMAPP, Radboud University Nijmegen, P.O. Box 9010, 6500 GL Nijmegen, The Netherlands }
\affiliation {Artemis, Universit\'e C\^ote d'Azur, CNRS, Observatoire C\^ote d'Azur, CS 34229, F-06304 Nice Cedex 4, France }
\affiliation {Institut de Physique de Rennes, CNRS, Universit\'e de Rennes 1, F-35042 Rennes, France }
\affiliation {Washington State University, Pullman, WA 99164, USA }
\affiliation {Universit\`a degli Studi di Urbino 'Carlo Bo', I-61029 Urbino, Italy }
\affiliation {INFN, Sezione di Firenze, I-50019 Sesto Fiorentino, Firenze, Italy }
\affiliation {University of Oregon, Eugene, OR 97403, USA }
\affiliation {Laboratoire Kastler Brossel, UPMC-Sorbonne Universit\'es, CNRS, ENS-PSL Research University, Coll\`ege de France, F-75005 Paris, France }
\affiliation {Carleton College, Northfield, MN 55057, USA }
\affiliation {Astronomical Observatory Warsaw University, 00-478 Warsaw, Poland }
\affiliation {VU University Amsterdam, 1081 HV Amsterdam, The Netherlands }
\affiliation {University of Maryland, College Park, MD 20742, USA }
\affiliation {Laboratoire des Mat\'eriaux Avanc\'es (LMA), CNRS/IN2P3, F-69622 Villeurbanne, France }
\affiliation {Universit\'e Claude Bernard Lyon 1, F-69622 Villeurbanne, France }
\affiliation {Universit\`a di Napoli 'Federico II', Complesso Universitario di Monte S.Angelo, I-80126 Napoli, Italy }
\affiliation {NASA/Goddard Space Flight Center, Greenbelt, MD 20771, USA }
\affiliation {RESCEU, University of Tokyo, Tokyo, 113-0033, Japan. }
\affiliation {University of Adelaide, Adelaide, South Australia 5005, Australia }
\affiliation {Tsinghua University, Beijing 100084, China }
\affiliation {Texas Tech University, Lubbock, TX 79409, USA }
\affiliation {The University of Mississippi, University, MS 38677, USA }
\affiliation {The Pennsylvania State University, University Park, PA 16802, USA }
\affiliation {National Tsing Hua University, Hsinchu City, 30013 Taiwan, Republic of China }
\affiliation {Charles Sturt University, Wagga Wagga, New South Wales 2678, Australia }
\affiliation {University of Chicago, Chicago, IL 60637, USA }
\affiliation {Kenyon College, Gambier, OH 43022, USA }
\affiliation {Korea Institute of Science and Technology Information, Daejeon 305-806, Korea }
\affiliation {University of Cambridge, Cambridge CB2 1TN, United Kingdom }
\affiliation {Universit\`a di Roma 'La Sapienza', I-00185 Roma, Italy }
\affiliation {University of Brussels, Brussels 1050, Belgium }
\affiliation {Sonoma State University, Rohnert Park, CA 94928, USA }
\affiliation {Montana State University, Bozeman, MT 59717, USA }
\affiliation {Center for Interdisciplinary Exploration \& Research in Astrophysics (CIERA), Northwestern University, Evanston, IL 60208, USA }
\affiliation {Universitat de les Illes Balears, IAC3---IEEC, E-07122 Palma de Mallorca, Spain }
\affiliation {The University of Texas Rio Grande Valley, Brownsville, TX 78520, USA }
\affiliation {Bellevue College, Bellevue, WA 98007, USA }
\affiliation {Institute for Plasma Research, Bhat, Gandhinagar 382428, India }
\affiliation {The University of Sheffield, Sheffield S10 2TN, United Kingdom }
\affiliation {California State University, Los Angeles, 5154 State University Dr, Los Angeles, CA 90032, USA }
\affiliation {Universit\`a di Trento, Dipartimento di Fisica, I-38123 Povo, Trento, Italy }
\affiliation {INFN, Trento Institute for Fundamental Physics and Applications, I-38123 Povo, Trento, Italy }
\affiliation {Cardiff University, Cardiff CF24 3AA, United Kingdom }
\affiliation {Montclair State University, Montclair, NJ 07043, USA }
\affiliation {National Astronomical Observatory of Japan, 2-21-1 Osawa, Mitaka, Tokyo 181-8588, Japan }
\affiliation {Canadian Institute for Theoretical Astrophysics, University of Toronto, Toronto, Ontario M5S 3H8, Canada }
\affiliation {MTA E\"otv\"os University, ``Lendulet'' Astrophysics Research Group, Budapest 1117, Hungary }
\affiliation {School of Mathematics, University of Edinburgh, Edinburgh EH9 3FD, United Kingdom }
\affiliation {University and Institute of Advanced Research, Gandhinagar, Gujarat 382007, India }
\affiliation {IISER-TVM, CET Campus, Trivandrum Kerala 695016, India }
\affiliation {University of Szeged, D\'om t\'er 9, Szeged 6720, Hungary }
\affiliation {International Centre for Theoretical Sciences, Tata Institute of Fundamental Research, Bengaluru 560089, India }
\affiliation {Embry-Riddle Aeronautical University, Prescott, AZ 86301, USA }
\affiliation {Tata Institute of Fundamental Research, Mumbai 400005, India }
\affiliation {INAF, Osservatorio Astronomico di Capodimonte, I-80131, Napoli, Italy }
\affiliation {University of Michigan, Ann Arbor, MI 48109, USA }
\affiliation {Rochester Institute of Technology, Rochester, NY 14623, USA }
\affiliation {NCSA, University of Illinois at Urbana-Champaign, Urbana, IL 61801, USA }
\affiliation {University of Bia{\l }ystok, 15-424 Bia{\l }ystok, Poland }
\affiliation {SUPA, University of Strathclyde, Glasgow G1 1XQ, United Kingdom }
\affiliation {University of Southampton, Southampton SO17 1BJ, United Kingdom }
\affiliation {University of Washington Bothell, 18115 Campus Way NE, Bothell, WA 98011, USA }
\affiliation {Institute of Applied Physics, Nizhny Novgorod, 603950, Russia }
\affiliation {Seoul National University, Seoul 151-742, Korea }
\affiliation {Inje University Gimhae, 621-749 South Gyeongsang, Korea }
\affiliation {National Institute for Mathematical Sciences, Daejeon 305-390, Korea }
\affiliation {Pusan National University, Busan 609-735, Korea }
\affiliation {NCBJ, 05-400 \'Swierk-Otwock, Poland }
\affiliation {Institute of Mathematics, Polish Academy of Sciences, 00656 Warsaw, Poland }
\affiliation {Monash University, Victoria 3800, Australia }
\affiliation {Hanyang University, Seoul 133-791, Korea }
\affiliation {The Chinese University of Hong Kong, Shatin, NT, Hong Kong }
\affiliation {University of Alabama in Huntsville, Huntsville, AL 35899, USA }
\affiliation {ESPCI, CNRS, F-75005 Paris, France }
\affiliation {University of Minnesota, Minneapolis, MN 55455, USA }
\affiliation {Universit\`a di Camerino, Dipartimento di Fisica, I-62032 Camerino, Italy }
\affiliation {Southern University and A\&M College, Baton Rouge, LA 70813, USA }
\affiliation {The University of Melbourne, Parkville, Victoria 3010, Australia }
\affiliation {College of William and Mary, Williamsburg, VA 23187, USA }
\affiliation {Instituto de F\'\i sica Te\'orica, University Estadual Paulista/ICTP South American Institute for Fundamental Research, S\~ao Paulo SP 01140-070, Brazil }
\affiliation {Whitman College, 345 Boyer Avenue, Walla Walla, WA 99362 USA }
\affiliation {Universit\'e de Lyon, F-69361 Lyon, France }
\affiliation {Hobart and William Smith Colleges, Geneva, NY 14456, USA }
\affiliation {Janusz Gil Institute of Astronomy, University of Zielona G\'ora, 65-265 Zielona G\'ora, Poland }
\affiliation {King's College London, University of London, London WC2R 2LS, United Kingdom }
\affiliation {IISER-Kolkata, Mohanpur, West Bengal 741252, India }
\affiliation {Indian Institute of Technology, Gandhinagar Ahmedabad Gujarat 382424, India }
\affiliation {Andrews University, Berrien Springs, MI 49104, USA }
\affiliation {Universit\`a di Siena, I-53100 Siena, Italy }
\affiliation {Trinity University, San Antonio, TX 78212, USA }
\affiliation {University of Washington, Seattle, WA 98195, USA }
\affiliation {Abilene Christian University, Abilene, TX 79699, USA }


\begin{abstract}

We present the results from an all-sky search for short-duration gravitational waves in the 
data of the first run of the Advanced LIGO detectors between September 2015 and January 2016.
The search algorithms use minimal assumptions on the signal morphology, so they are 
sensitive to a wide range of sources emitting gravitational waves. The analyses target transient signals
with duration ranging from milliseconds to seconds over the frequency band of 32 to 4096 Hz.
The first observed gravitational-wave event, GW150914, has been detected with high confidence in this search; other 
known gravitational-wave events fall below the search's sensitivity.
Besides GW150914, all of the search results are consistent with the expected rate of accidental noise coincidences.  
Finally, we estimate
rate-density limits for a broad range of non-BBH transient gravitational-wave sources as a function of their gravitational radiation 
emission energy and their characteristic frequency.  
These rate-density upper-limits are stricter than those previously published by an order-of-magnitude.

\end{abstract}

\maketitle

\section{Introduction}

The first observing period of the Advanced LIGO detectors \cite{TheLIGOScientific:2014jea, TheVirgo:2014hva}
has been completed recently with the most sensitive
gravitational-wave (GW) detectors ever built. The two LIGO observatories in Hanford, WA and Livingston, LA
achieved a major milestone in gravitational wave astronomy: the first direct detection
of gravitational waves on September 14, 2015, referred as GW150914 \cite{DetectionPaper}.
Advanced LIGO is the first of a new generation of instruments, including GEO 600 \cite{Luck:2010rt}, Advanced Virgo \cite{TheVirgo:2014hva}, 
KAGRA \cite{Aso:2013eba} and LIGO-India \cite{2013arXiv1304.0670L}.

This paper reports on a search for short-duration transient gravitational-wave events, commonly referred to as GW bursts, 
during the first observing run (O1) of the Advanced LIGO detectors, from September 2015 to January 2016. 
The first 16 days of coincident data have been already analyzed, resulting in a high-significance detection statement for the GW150914
event \cite{BurstCompanion}.
GW bursts can be generated by a wide variety of astrophysical sources, such as merging compact binary systems \cite{Aasi:2014iwa, Mohapatra:2014rda}, 
core-collapse supernovae of massive stars \cite{Fryer:2011zz}, 
neutron stars collapsing to form black holes, pulsar glitches, and cosmic string cusps \cite{PhysRevD.71.063510}. 
To search broadly for these phenomena, 
we employ searches with minimal assumptions
regarding the expected waveform characteristics and the source direction. 
The search we report here is more sensitive than the previous burst searches \cite{s6Burst}
because of both the increased sensitivity of the Advanced detectors \cite{Aasi:2014mqd} and improvements in the search algorithms in rejecting 
transient non-Gaussian noise artifacts (glitches) \cite{Kanner:2015xua, Littenberg:2015kpb, oLIBprep, cwb2g}.

The described un-modeled all-sky search for GW bursts consists of three different algorithms. 
This paper shows the result of these algorithms, and gives limits on the rate-density of transient GW events.
All of these algorithms have independently claimed high-significance detections of GW150914 \cite{BurstCompanion}. 
The lower-mass GW event, GW151226 \cite{BoxingDay}, and the LVT151012 candidate \cite{CBC_Companion,CBC_O1} were not detected by these searches. 

The paper is organized as follows: in Section \ref{Sec: o1run} we give an overview of the O1 data set. 
In Section \ref{Sec: searches} we give a brief overview of the three search algorithms. 
The sensitivity of the search is described in Section \ref{Sec: sensitivity}.
Finally, Sections \ref{Sec: results} and \ref{Sec: discussion} discuss the search results and their implications.

\section{Observing Run 1}\label{Sec: o1run}

Our data set extends over 130 calendar days from September 12, 2015 to January 19, 2016. This first observing period (called O1)
of Advanced LIGO began after a series of major upgrades to both the Hanford and Livingston detectors \cite{DetectionPaper}.

  
In the most sensitive frequency band, 100-300 Hz, the O1 LIGO detectors are 3 to 5 times more sensitive than the initial 
LIGO detectors \cite{Aasi:2014mqd}. Future observing runs are expected to increase sensitivity by an additional factor of 3 \cite{2013arXiv1304.0670L}.


As in the previous LIGO/Virgo searches \cite{Abbott:2015vir, Abadie:2012rq, Abadie:2010mt}, intervals of poor data quality are identified and excluded from the analysis.
To monitor environmental disturbances and their influence on the detectors, each observatory is equipped with an array of sensors: seismometers, accelerometers, microphones, magnetometers, radio receivers, weather sensors, ac-power line monitors, and a cosmic-ray detector. 
Hundreds of thousands of auxiliary channels within the instrument are also monitored. 
Characterization of the relationship of the strain data to this additional information allows many non-GW transients to be removed with high statistical confidence \cite{DetcharCompanion, BurstCompanion}.

The livetime in which the two detectors were individually locked is about 79 days for H1 and 67 days for L1. 
After data quality flags have been applied, the total analyzable
time is about 75 days for H1 and 65 days for L1. The coincident livetime between H1 and L1 is about 48 days.
This livetime includes the 16 days of this coincident data that has already been analyzed in \cite{BurstCompanion}.
Finally, the estimated calibration uncertainty ($1\sigma$) below 2 kHz is less than
10\% in amplitude and 10 degrees in phase \cite{Calibration}. The calibration uncertainty above 2 kHz is less certain, although the limited data 
obtained at these frequencies suggests upper bounds of 20\% in amplitude and 10 degrees in phase. These estimates will be 
further refined through future measurements and analyses \cite{CalibrationHF}.

\section{Searches}\label{Sec: searches}

This search covers the most sensitive frequency band of the involved detectors, i.e. 32 - 4096 Hz, 
and it consists of the same three burst algorithms used to measure 
the significance of GW150914 \cite{BurstCompanion}.
They consist of two end-to-end algorithms: coherent Waveburst (cWB) \cite{klimenko:2008fu,cwb2g}
and omicron-LIB (oLIB) \cite{oLIBprep}; and a follow-up algorithm applied to cWB events:
BayesWave (BW) \cite{Cornish:2014kda,Littenberg:2014oda}.
Using multiple search algorithms has two advantages: it can provide independent validation of results, and it can also improve the search sensitivity in regions of parameter space where a single algorithm outperforms the others.

The three algorithms ran over the 48 days of coincident data. However, due to internal segmentation\footnote{The cWB algorithm requires at least 600 s of continuous data to perform its analysis.}
the cWB and BW pipelines only actually analyzed 44 days of this coincident data.
The oLIB analysis loss-time is negligible and thus oLIB analyzed close to the full 48 days.

The three algorithms also ran in low-latency mode during O1.
In this mode, both cWB and oLIB produced independent alerts
of the GW150914 event and the result was validated by a BW follow-up \cite{EMCompanion}.

To characterize the statistical rate of transient noise glitches occurring simultaneously at the two LIGO sites by chance,
this analysis uses the time-shift method: data from one interferometer is shifted in time with respect to the other interferometer 
by multiple delays much larger than the maximum GW travel time between the interferometers. 
In this way, we can accumulate a significant duration of estimated background that we use 
to estimate the false-alarm rate (FAR) for each algorithm.


We set a FAR threshold of 1 in 100 years for identifying a detection candidate, which roughly corresponds to a 3 sigma detection statement for the duration of our observation. 
If an event in this search were to have a FAR less than this threshold, a refined analysis (i.e., more time-shifts) would be performed to assign the appropriate significance in the detection statement for this event.

\begin{figure*}[!htb]
 \begin{center}
  \subfigure[\textbf{cWB} 32-1024 Hz search classes: $C1$ (red), $C2$ (brown), $C3$ (yellow).\label{Fig. cWBLFzerolag}]{\includegraphics[width=250px]{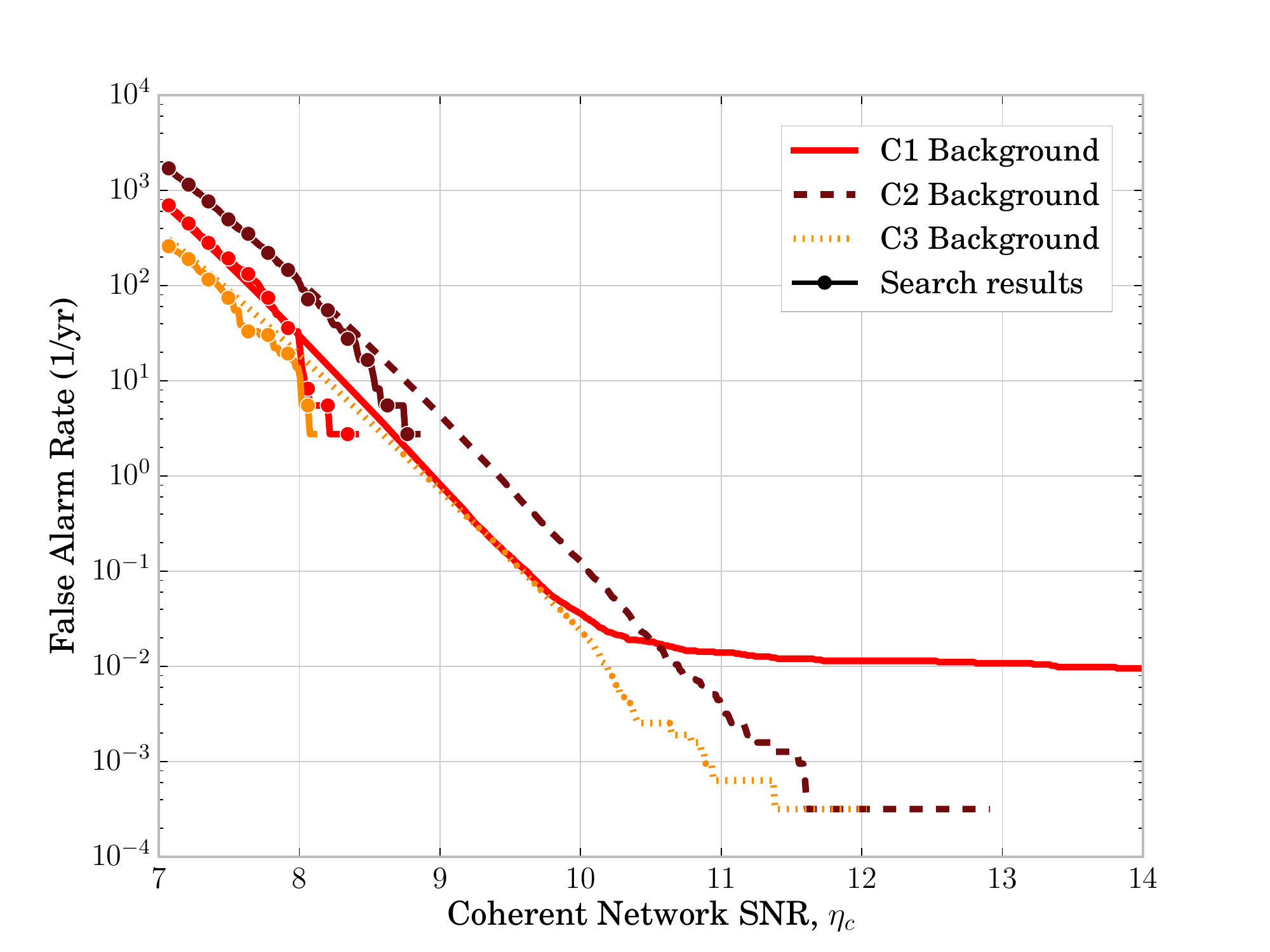}}
  \subfigure[\textbf{cWB} 1024-4096 Hz search class.\label{Fig. cWBHFzerolag}]{\includegraphics[width=250px]{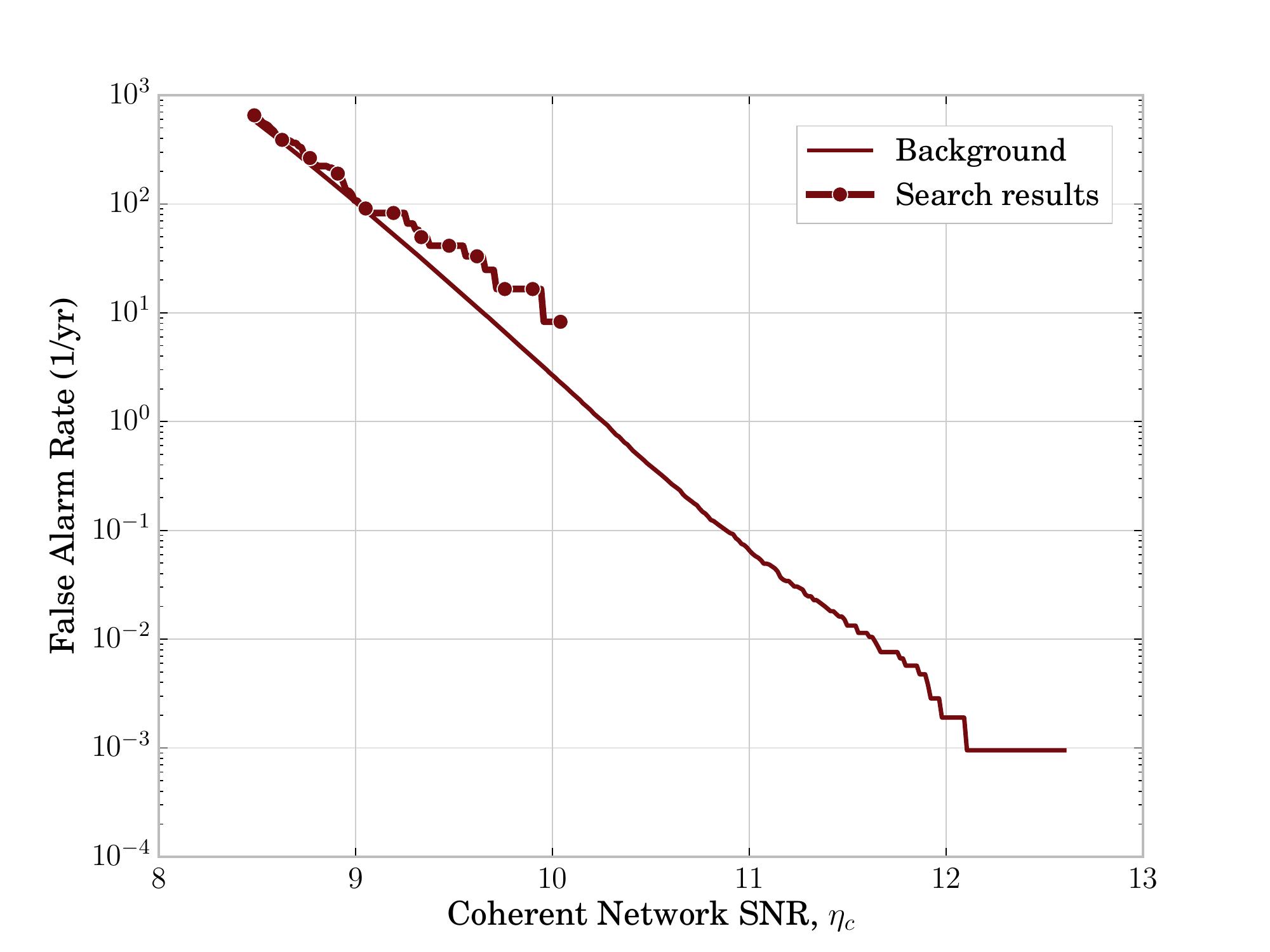}}\\
  \subfigure[\textbf{oLIB} 48-1024 Hz low-$Q$ (dashed) and high-$Q$ (solid)  search classes.\label{Fig.oLIB_FAR}]{\includegraphics[width=250px]{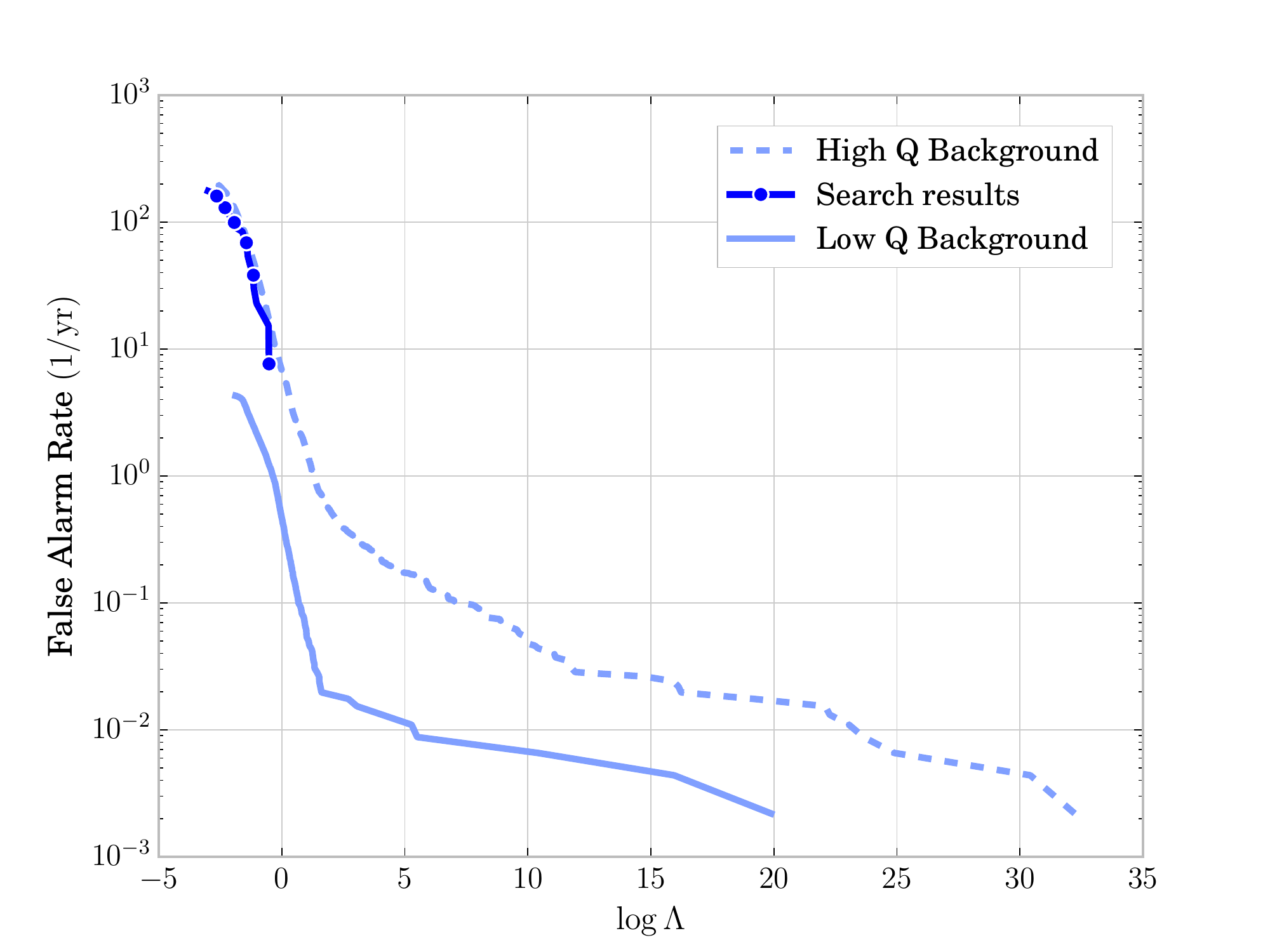}}
  \subfigure[\textbf{BayesWave} followup to cWB 32-1024 Hz search class.\label{Fig.BW_bg}]{\includegraphics[width=250px]{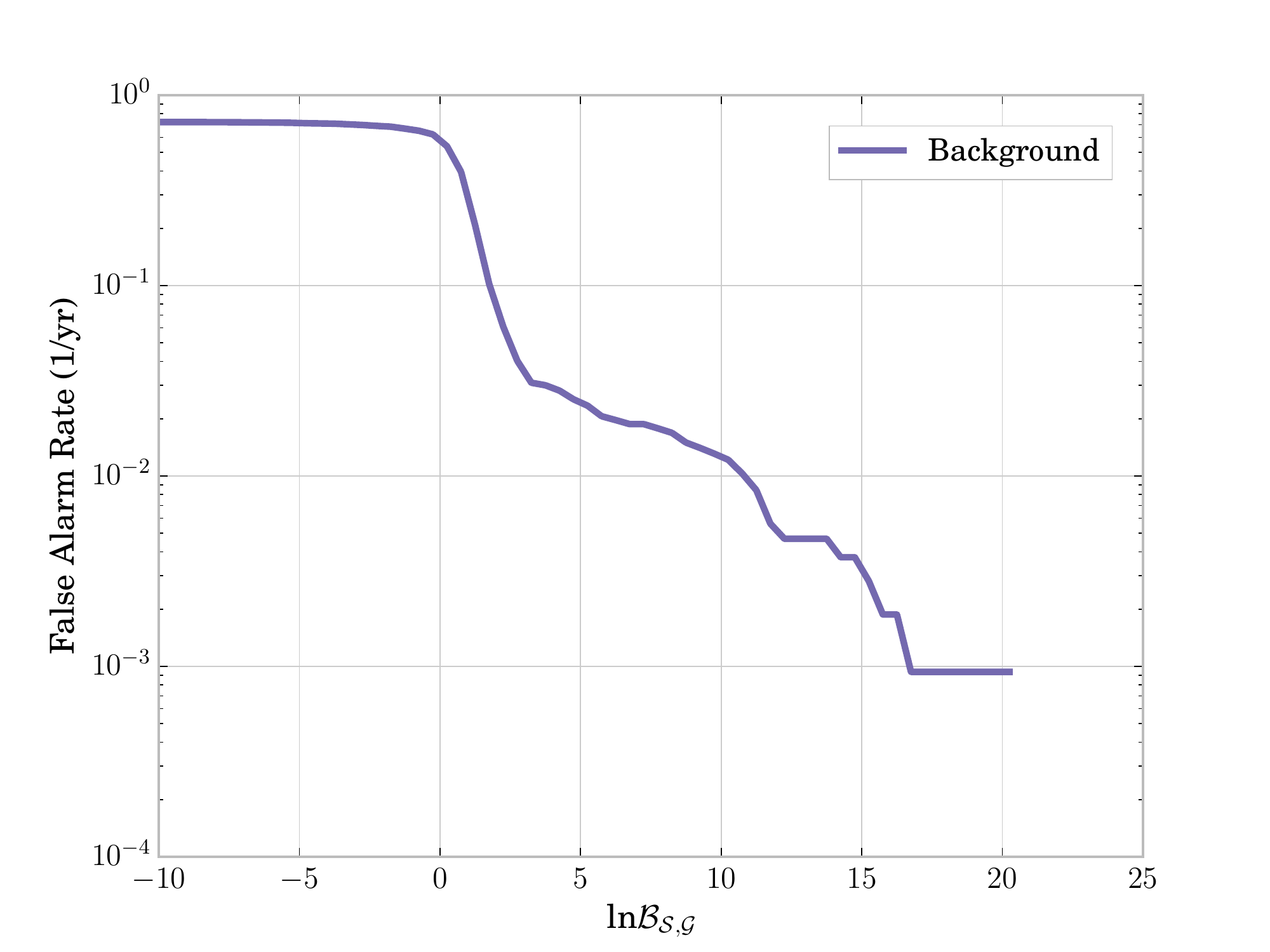}}
 \end{center}
\caption{Search results and backgrounds as a function of the detection statistic for the different searches.  The FAR refers to the rate at which events more significant than the corresponding detection statistic occur. Apart from GW150914 
(which is not reported in these figures), the search results are consistent with the expectations of accidental noise coincidences.}
\end{figure*}

\subsection{Coherent WaveBurst}

Coherent WaveBurst (cWB) has been used in multiple searches for transient GWs \cite{Abadie:2010mt, s6Burst}. 
It calculates a maximum-likelihood-ratio statistic for power excesses identified in the time-frequency domain.
A primary selection cut
is applied to the network correlation coefficient $c_c$, which measures the degree of correlation between the detectors. 
Events with $c_c<0.7$ are discarded from the analysis. Events are ranked according to their
coherent network signal-to-noise ratio (SNR) $\eta_c$, which is related to the matched-filter SNR, favoring GW signals correlated in both detectors and suppressing uncorrelated glitches. A detailed explanation of the algorithm
and the definition of these statistics are given in \cite{BurstCompanion}.

The cWB analysis is divided in two frequency bands, where the splitting frequency is 1024 Hz. For the low-frequency band,
the data is downsampled to reduce the computational cost of the analysis. 

Low-frequency cWB events are divided into three search classes according to their morphology, as described in \cite{BurstCompanion}. 
The $C1$ class is based on cuts which primarily select so-called ``blip'' glitches and non-stationary power-spectrum lines. 
The former are non-Gaussian noise transients of unknown origin consisting of a few cycles around 100 Hz.
The $C3$ class is based on cuts that select events whose frequency increases with time, i.e. those similar in morphology to the merger of compact objects. 
The $C2$ class is composed of all the remaining events.

The FAR of each identified event is estimated using the time-slide background distribution of similar class. Since there are three independent classes, we apply a trials factor of 3 to estimate the final significance.  The high-frequency analysis consists of only a single class.



About 1000 years of coincident background data were accumulated for the cWB analysis.
Fig. \ref{Fig. cWBLFzerolag} and \ref{Fig. cWBHFzerolag} report the cumulative FAR as a function of $\eta_c$ for the 
low-frequency and high-frequency analyses, respectively, including the three different classes for the low-frequency case.  


\subsection{Omicron-LIB}

Omicron-LIB (oLIB) is a hierarchical search algorithm that first analyzes the data streams of individual detectors, which we refer to as an incoherent analysis.  It then follows up stretches of data that are potentially correlated across the detector network, which we refer to as a coherent analysis.
The incoherent analysis (``Omicron'')~\cite{OmicronRobinet} flags stretches of coincident excess power.
The coherent follow-up (``LIB'')~\cite{oLIBprep} models gravitational wave signals and noise transients with a single sine-Gaussian, and it produces two different Bayes factors.
Each of these Bayes factors is expressed as the natural logarithm of the evidence ratio of two hypotheses: a GW signal \textit{vs} Gaussian noise (BSN) and a coherent GW signal \textit{vs} incoherent noise transients (BCI).
The joint likelihood ratio $\Lambda$ of these two Bayes factors is used as ranking statistic to assign a significance to each event.
See~\cite{oLIBprep} for further technical details on the implementation of these steps.

For this analysis, oLIB events are divided into two classes, based on the inferred parameters of the best-fit sine-Gaussian.
The exact parameter ranges of these search classes are chosen in order to group noise transients of similar morphology together.
Particularly noisy regions of the parameter space are excluded from the analysis entirely (e.g., events with median quality factor $Q$ > 108).
Both classes contain only events whose median frequency $f_0$, as estimated by LIB, lies within the range of 48 - 1024 Hz. 
The first, analogous to cWB's C1 class, is a ``low-Q'' class that contains only events whose median quality factor $Q$, lies within the range 0.1 - 2. 
The second, analogous to the union of cWB's C2 and C3 classes, is a ``high-Q'' class that contains only events whose median $Q$ lies within the range 2 - 108. 
In both classes, event candidates were also required to have positive Bayes factors, i.e., BSN > 0 and BCI > 0, meaning the evidence for the signal model was greater than the evidences for the noise models.
A trials factor of 2 accounts for these independent search classes.


The oLIB background analysis is performed using 456 years of background data.
We select single-detector events with SNR > 5.0. 
This is lower than the threshold of 6.5 adopted in \cite{BurstCompanion} and it is chosen to allow us to make a significance estimation of low-SNR events. 
For this reason, we cannot directly compare the two set of results reported in \cite{BurstCompanion} and in this study using the likelihood ratios $\Lambda$, but we have to consider the reported FAR. 
The results are presented in Fig.~\ref{Fig.oLIB_FAR}.




\subsection{BayesWave Follow-up}

BayesWave (BW) tests if the data in multiple detectors are best explained by coincident glitches or a signal, and it is used as a follow-up to events produced by cWB.
It has been shown that BW is able to increase the detection confidence for GW signals of complex morphology\cite{Kanner:2015xua}.  

The BW algorithm uses a variable number of sine-Gaussian wavelets to reconstruct the data independently for the signal and glitch models, then computes the natural logarithm of the Bayes factor between these two models, $\ln\mathcal{B}_{sg}$.  The number of wavelets used is determined by using a reversible jump Markov chain Monte Carlo, with more complex signals requiring more wavelets \cite{Littenberg:2015kpb}.  The Bayes factor scales as $\ln{\mathcal{B}_{sg}}\sim N \ln \mathrm{SNR}$, where $N$ is number of wavelets used. This means the detection statistic depends on waveform complexity in addition to SNR.  Full details of the algorithm can be found in \cite{Cornish:2014kda}.

In this search, BW followed up events produced by cWB in any of the three low-frequency search classes
with a coherent network SNR of $\eta_c \geq 9.9$
and correlation coefficient of $c_c>0.7$.  
There are no additional cuts performed on the data, and all of these events (C1+C2+C3) are analyzed as a single class. 
The cumulative FAR as a function of  $\ln\mathcal{B}_{sg}$ is shown in Fig. \ref{Fig.BW_bg}. 


\section{Sensitivity} \label{Sec: sensitivity}

The detection efficiency of the search is measured by adding simulated signals into the detectors' data and
evaluating whether or not they pass the selection cuts explained in the Section \ref{Sec: searches} for the different search algorithms.
A variety of GW signal morphologies were tested, spanning a wide range of amplitudes and duration, and with
characteristic frequencies within the sensitive bandwidth of the detectors. We identify two different waveform
sets: 
a set of generic bursts,
and a set of simulated astrophysical signals coming from the coalescence and merging of binary black holes (BBH).
All of the results in this section refer to a FAR detection threshold of 1 in 100 years.

\subsection{Generic bursts}

This family includes the waveform types described in \cite{Abadie:2012rq}, all with elliptical polarization:
\textit{gaussian pulses (GA)}, parametrized by their duration parameter $\tau$; \textit{sine-Gaussian wavelets (SG)},
sinusoids within a Gaussian envelope, characterized by the frequency of the sinusoid $f_0$ and
a quality factor $Q$;
\textit{white-noise bursts (WNB)}, white noise bounded in frequency over a bandwidth $\Delta f$ and with a Gaussian envelope, described by the lower frequency 
$f_{low}$, $\Delta f$, and the duration $\tau$. 
Table \ref{Tab.hrss50} lists the waveforms that have been considered for this work.


The amplitudes of the test signals are chosen to cover a wide range of values and are expressed in terms of the root-mean-square
strain amplitude at Earth (before accounting for the detection response patterns), denoted $h_{\text{rss}}$ \cite{s6Burst}.
For this search, we injected signals according to the distance distribution $p(r)=r+A/r$ where $A$ is a constant. 
The constant $A$ is chosen to produce at least several test events with large $h_\text{rss}$.




\begin{table}[!htb]
\centering
\scriptsize
\begin{tabular}{|c||c|c|c|}
\hline
\textbf{Morphology} & \textbf{cWB} & \textbf{oLIB} & \textbf{BW} \\
\hline
\textbf{Gaussian pulses}\\
\hline
$\tau = 0.1$ ms & 34 & N/A & N/A \\
\hline
$\tau = 2.5$ ms & 33 & 7.4 & N/A \\
\hline
\textbf{sine-Gaussian wavelets}\\
\hline
$f_0 = 70$ Hz, $Q = 100$ & 24 & N/A & N/A \\
\hline
$f_0 = 153$ Hz, $Q = 8.9$ & 1.6 & 1.7 & 5.4 \\
\hline
$f_0 = 235$ Hz, $Q = 100$ & 14 & 1.9 & N/A \\
\hline
$f_0 = 554$ Hz, $Q = 8.9$ & 2.6 & 2.7 & 3.6 \\
\hline
$f_0 = 849$ Hz, $Q = 3$ & 27 & 3.3 & 5.4 \\
\hline
$f_0 = 1615$ Hz, $Q = 100$ & 5.5 & - & - \\
\hline
$f_0 = 2000$ Hz, $Q = 3$ & 8.7 & - & - \\
\hline
$f_0 = 2477$ Hz, $Q = 8.9$ & 11 & - & - \\
\hline
$f_0 = 3067$ Hz, $Q = 3$ & 15 & - & - \\
\hline
\textbf{White-Noise Bursts}\\
\hline
$f_{low} = 100$ Hz, $\Delta f = 100$ Hz, $\tau = 0.1$ s & 2.0 & N/A & 3.0 \\
\hline
$f_{low} = 250$ Hz, $\Delta f = 100$ Hz, $\tau = 0.1$ s & 2.2 & N/A & 9.2 \\
\hline
\end{tabular}
\caption{The $h_\text{rss}$ values, in units of $10^{-22} \text{Hz}^{-1/2}$, at which 50\% detection efficiency is achieved at a FAR of 1 in 100 yr for each of the algorithms, as a function of the injected signal morphologies.  ``N/A'' denotes that 50\% detection efficiency was not achieved. ``-'' denotes the waveform was not analyzed by oLIB and BW because its characteristic frequency is higher than 1024 Hz.}\label{Tab.hrss50}
\end{table}

Table \ref{Tab.hrss50}, shows the $h_\text{rss}$ value at which 50\% of the injections are detected for each signal morphology and algorithm.
There are some morphology-dependent features that affect each of the different algorithms at the FAR threshold of 1 in 100 years.
These features largely disappear and the different algorithms' results converge at detection thesholds of higher FAR.
For example, the detection efficiencies are worse for cWB for low-Q morphologies and high-Q morphologies because these injections are
classified as C1 events.  
As shown in Fig. \ref{Fig. cWBLFzerolag}, the C1 background extends to higher significances than in the other bins, meaning these high-Q and low-Q events must have large values of $\eta_c$ to meet the FAR threshold of 1 in 100 years.
The oLIB detection efficiencies, while non-negligible across all morphologies, never quite reach 50\% for some non-sine-Gaussian morphologies because
the template mismatch residuals grow linearly with $h_\text{rss}$.
Finally, the detection efficiencies of BW suffers for high-Q events since its prior range only extends to $Q=40$.
However, almost every morphology can be detected efficiently by at least one of the algorithms.

\begin{figure}
    \includegraphics[width=250px]{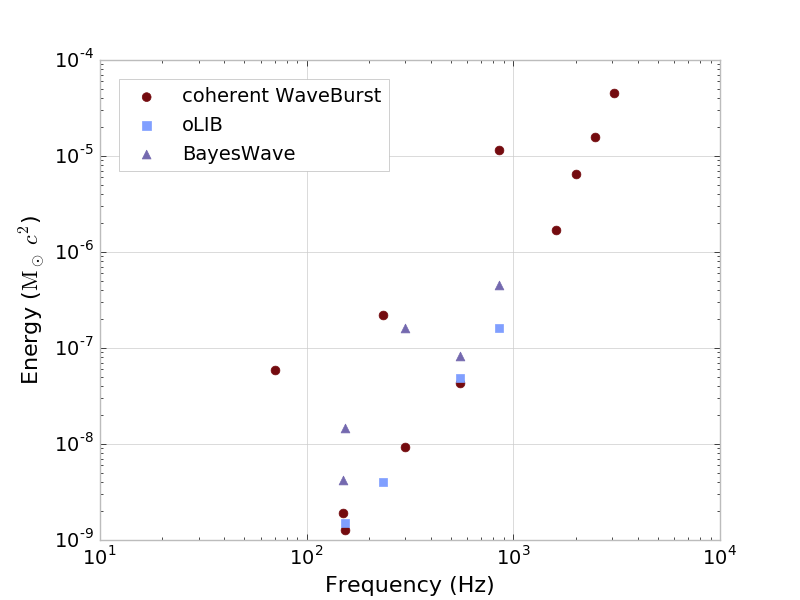}
    \caption{GW emission energy, in solar masses, at 50\% detection efficiency for standard-candle sources emitting at 10 kpc for
the non-GA waveforms listed in Table \ref{Tab.hrss50}.  These results can be scaled to any reference distance $r_0$ using $E_{\text{GW}} \propto r_0^2$.}
    \label{Fig. cWB E vs f}
\end{figure}

Another way to interpret the search sensitivities is to map them into the minimum amount of energy that needs to be emitted through GWs for at least half of the sources to be detected within a given search volume.
Assuming a fixed amount of energy is radiated isotropically away from the source in GWs of a fixed frequency $f_0$, this distance $r_0$ can
be converted into a value of $h_{\text{rss}}$ via the relationship \cite{s6Burst}:
  \begin{equation}\label{Eq.E_to_f}
  E_{\text{GW}} = \frac{\pi^2 c^3}{G}r_0^2f_0^2h_0^2
  \end{equation}
Here, we use the $h_{\text{rss}}$ from Table \ref{Tab.hrss50}, the central frequency of each morphology, and a fixed fiducial radius to calculate this energy via Eq.~\ref{Eq.E_to_f}.  Figure~\ref{Fig. cWB E vs f} shows this energy as a function of characteristic frequency assuming a galactic source at a distance of 10 kpc.
When taking into account the results of all three algorithms, this emission energy is not strongly dependent on the type of waveform (with exceptions on an algorithm-by-algorithm basis, as described above).
Fig.~\ref{Fig. cWB E vs f} can easily be converted to other distances by applying the scaling relation suggested by Eq.~\ref{Eq.E_to_f}.
Previous studies \cite{s6Burst} have published similar emission-energy-versus-frequency plots at a detection threshold of 1 in 8 years.
We note that the current results, when evaluated at this higher-FAR threshold, are roughly an order-of-magnitude more sensitive than these previous results,
due mainly to the improvement in detector sensitivites. 

\subsection{Binary black holes mergers}

We also consider a set of astrophysical waveforms using models of merging of binary black hole systems.
Specifically, we choose the SEOBNRv2 model as implemented in the LAL software library \cite{Taracchini:2013rva,Kumar:2015tha}.
The waveforms are generated with an initial frequency of $15$ Hz.
The simulated binary systems are isotropically located in the sky and isotropically oriented. 
The total redshifted mass of the system in the detector frame\footnote{Given the luminosity distance of the system, one can assume a cosmology and calculate its redshift $z$.  The system's total mass in the source frame can then be obtained by dividing the total redshifted mass in the detector frame by $(1+z)$.} is distributed uniformly between 10 and 150 $M_\odot$,
a range that encompasses the total masses of both GW150914 and GW151226 \cite{CBC_O1}. The black hole spins are aligned with the binary angular momentum, and the magnitude of the dimensionless spin vector, $\textbf{a}_{1,2}$, is uniformly distributed between 0 and 0.99.
We neglect any cosmological corrections, such as normalizing our spatial distibution to be constant in co-moving volume.
We generate three different injection sets, each one with a mass ratio $q = m_2/m_1$ chosen from the set $\{0.25,0.5,1.0\}$ (where $m_1$ is by definition the more massive object). 

\begin{figure*}
    \includegraphics[width=500px]{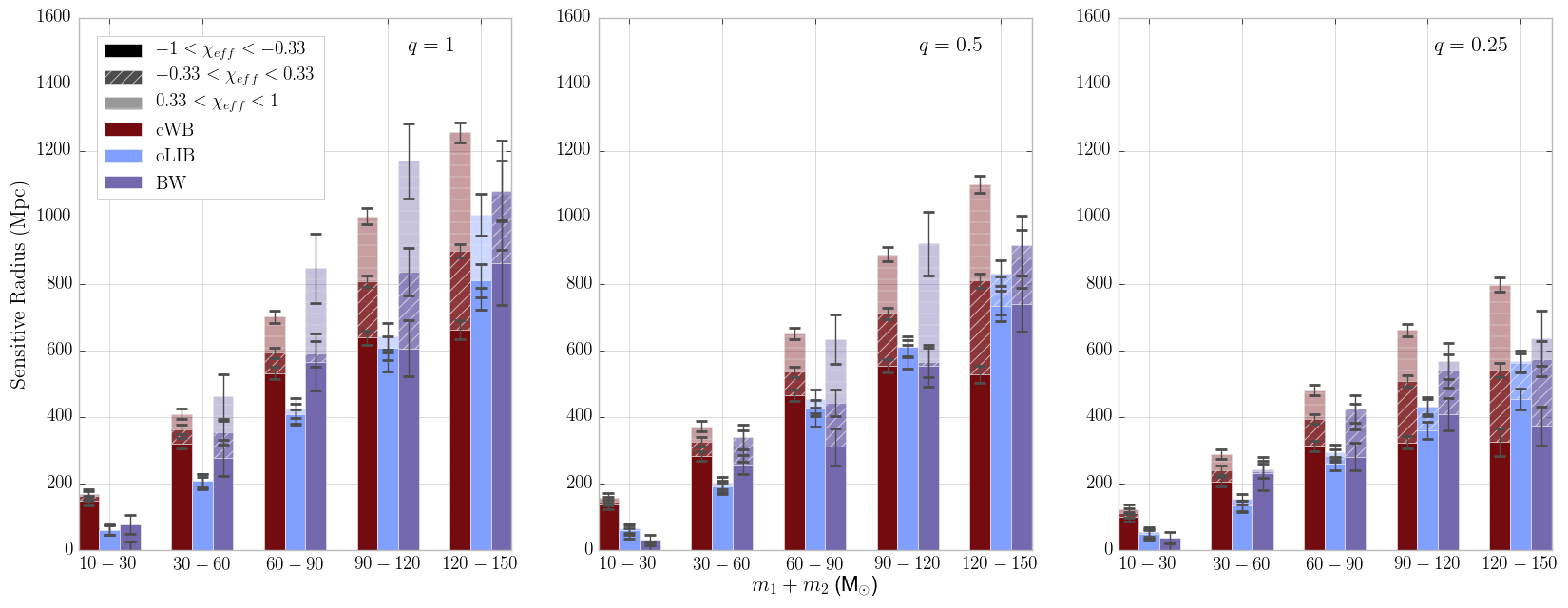}
    \caption{A comparison of the sensitive luminosity radii \cite{BurstCompanion} in Mpc, as a function of the total redshifted masses in the detector frame, among the three algorithms.
    The radii are binned according to mass ratio q (from left to right $q=1,0.5,0.25$) and effective spin $\chi_{\text{eff}}$, defined in \cite{BurstCompanion}. The three ranges of spin refer to aligned 
($0.33 < \chi_{eff} < 1$), non-spinning ($-0.33 < \chi_{eff} < 0.33$), anti-aligned ($-1 < \chi_{eff} < -0.33$).}
    \label{Fig. BBH dist}
\end{figure*}

In Fig.~\ref{Fig. BBH dist} we compare the sensitive luminosity radius \cite{2012arXiv1201.5999T} as a function of the total redshifted mass in the detector frame.
While systems inside this distance may be missed and systems outside of it may be detected depending on their sky position and orientation, this sensitive radius provides a ``rule-of-thumb'' determination on whether or not this burst search will detect a system's GW transients.
We can see that for systems like GW150914 ($\sim 70 M_\odot$ \cite{PE}) and GW151226 ($\sim 20 M_\odot$ \cite{CBC_O1}), the search ranges at the FAR of 1/100 years are approximately 500-700 Mpc and 100-200 Mpc, respectively.
These ranges demonstrates why this search detects GW150914 ($\sim 400$ Mpc \cite{PE}) but not GW151226 ($\sim 400$ Mpc \cite{CBC_O1}).
Even though the two sources are at a similar luminosity distance, this burst search is less efficient at detecting low-mass BBH systems.  
This behavior is true for two reasons:  lower-mass systems emit less energy into GWs than higher-mass systems, and this energy is distributed over a longer duration of time. 
These two features make it more difficult for non-templated algorithms to extract the GW signal from the detector noise as compared to searches based on templates.


\section{Results}\label{Sec: results}

The most significant event and only detection established in this search is GW150914 \cite{DetectionPaper},
which is independently confirmed by all three algorithms.  Specifically, it is found by cWB in the C3 class of
the low-frequency analysis with an estimated FAR of less than 1 in 350 years, by oLIB in the 
``high-Q'' class with an estimated FAR of less than 1 in 230 years, and by BayesWave with an estimated FAR
of less than 1 in 1000 years.\footnote{Because GW150914 was louder than any of the background events 
in this search, we can only provide the relatively un-precise upper-limits on FAR listed above.} These 
results are less precise but consistent with \cite{DetectionPaper}.

All other events generated by the analyses are consistent with the accidental noise coincidence rates.
To be specific, there are no other events found above the SNR thresholds in either the ``low-Q'' class of oLIB or the entire
BayesWave analysis bin.  The rate of other events in the oLIB ``high-Q'' bin are consistent with the accidental
noise coincidence rates within 1 sigma.  The event in the cWB analysis with the second-lowest FAR 
belongs to the high frequency search, with a false-alarm probability of about 0.2. 

These results set constraints on the population of transient GW sources within the volume of the Universe that 
the detectors were sensitive to during O1.  Again, all of the results in this section refer to a FAR detection threshold of 1 in 100 years.

\begin{figure}
    \includegraphics[width=\columnwidth]{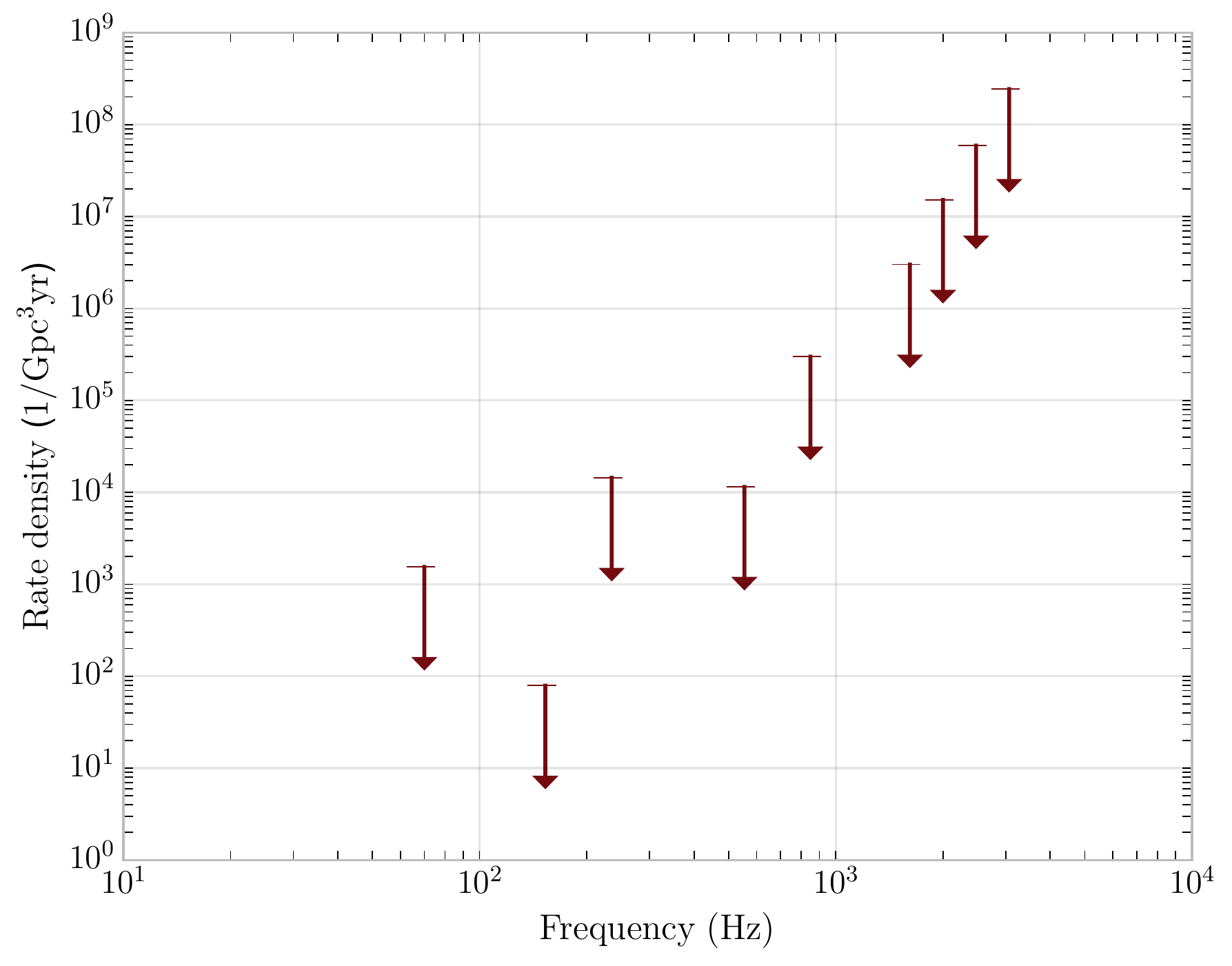}
    \caption{The 90\% confidence intervals of rate density given by the cWB pipeline
    for the sine-Gaussian waveforms listed in Table \ref{Tab.hrss50}.  
    This plot assumes zero detections, zero background, 
    and that 1 $M_\odot c^2$ of energy is emitted in gravitational waves.
    These results can be scaled to any emission energy $E_{GW}$ using $\textit{rate density} \propto E_{GW}^{-\frac{3}{2}}$.
    The arrow markers signify that the confidence intervals extend to zero.
    }
    \label{Fig. Upper limit}
\end{figure}

We estimate the limits on the rate density of generic non-BBH-like GW-burst sources in Fig. \ref{Fig. Upper limit} by removing the known BBH detections GW150914 and GW151226 from our analysis.
We use the sine-Gaussian injection set as a representative morphology, and present our cWB rate-density estimates as a function of their
characteristic frequencies.
The bands represent the 90\% confidence intervals on rate density \cite{s6Burst}, calculated using the Feldman-Cousins formalism for 0 background events \cite{FeldmanCousins}.
The frequency-dependent variation among the upper-limits is due to the sine-Gaussians falling into different cWB search classes
as a result of their specific value of $Q$.
For a given value of $Q$, the results follow a smoother frequency dependence.
These results are not directly comparable with those from previous
runs \cite{s6Burst} because of the different FAR detection thresholds.
However, we note that at the previously-used FAR detection threshold of 1 in 8 years, our search lowers these upper-limits by about an order of magnitude
across all frequencies.
The sensitivity improvements of the detectors and pipelines allow us
to make these stricter rate statements even though we analyzed less livetime compared
to \cite{s6Burst} (less than 50 days compared to 1.7 yr).
Fig.~\ref{Fig. Upper limit} assumes $1\ M_\odot c^2$ of gravitational wave energy has been emitted from the source, but this can be scaled to any emission energy $E_{GW}$ by using Eq.~\ref{Eq.E_to_f}. Note that the rate density scales as $\propto E_{GW}^{-\frac{3}{2}}$.

\section{Discussion}\label{Sec: discussion}

This paper reports the results for the search for short duration GW in the first Advanced LIGO observing run, with minimal 
assumptions on the signal waveform, direction or arrival time.
The two LIGO detectors, Livingston and Hanford, were operating from mid-September 2015 to mid-January 2016,
with a greater sensitivity to GWs than any previous LIGO-Virgo run. This search has been
performed considering two end-to-end algorithms and a follow-up algorithm.

The only detection established in this search is the GW150914 event, a binary system consisting of two black holes merging
to form a single one \cite{DetectionPaper}. The other known black hole detection \cite{BoxingDay} falls below the sensitivity of this search, and all other events in the search result are consistent with accidental noise coincidences between the detectors.



We report the minimum GW emission energy needed to detect at least half of the transient events emitted within some fiducial distance. These energies depend primarily on the signal frequency and are approximately constant over the 
different models of GW emission morphology. 
We also estimate rate-density limits on non-BBH transient sources as a function of their frequency and their gravitational wave emission energy.

The interferometric detectors LIGO and Virgo are currently being upgraded for the next scientific run. 
LIGO should improve its sensitivity over the next few years, Virgo should soon come online, and the implementation of KAGRA and
LIGO India is also in progress. All of these improvements will allow this type of un-modeled search to achieve a better sensitivity in the future \cite{2013arXiv1304.0670L}.

\bigskip\noindent\textit{Acknowledgments} ---
The authors gratefully acknowledge the support of the United States
National Science Foundation (NSF) for the construction and operation of the
LIGO Laboratory and Advanced LIGO as well as the Science and Technology Facilities Council (STFC) of the
United Kingdom, the Max-Planck-Society (MPS), and the State of
Niedersachsen/Germany for support of the construction of Advanced LIGO 
and construction and operation of the GEO600 detector. 
Additional support for Advanced LIGO was provided by the Australian Research Council.
The authors gratefully acknowledge the Italian Istituto Nazionale di Fisica Nucleare (INFN),  
the French Centre National de la Recherche Scientifique (CNRS) and
the Foundation for Fundamental Research on Matter supported by the Netherlands Organisation for Scientific Research, 
for the construction and operation of the Virgo detector
and the creation and support  of the EGO consortium. 
The authors also gratefully acknowledge research support from these agencies as well as by 
the Council of Scientific and Industrial Research of India, 
Department of Science and Technology, India,
Science \& Engineering Research Board (SERB), India,
Ministry of Human Resource Development, India,
the Spanish Ministerio de Econom\'ia y Competitividad,
the Conselleria d'Economia i Competitivitat and Conselleria d'Educaci\'o, Cultura i Universitats of the Govern de les Illes Balears,
the National Science Centre of Poland,
the European Commission,
the Royal Society, 
the Scottish Funding Council, 
the Scottish Universities Physics Alliance, 
the Hungarian Scientific Research Fund (OTKA),
the Lyon Institute of Origins (LIO),
the National Research Foundation of Korea,
Industry Canada and the Province of Ontario through the Ministry of Economic Development and Innovation, 
the Natural Science and Engineering Research Council Canada,
Canadian Institute for Advanced Research,
the Brazilian Ministry of Science, Technology, and Innovation,
Funda\c{c}\~ao de Amparo \`a Pesquisa do Estado de S\~ao Paulo (FAPESP),
Russian Foundation for Basic Research,
the Leverhulme Trust, 
the Research Corporation, 
Ministry of Science and Technology (MOST), Taiwan
and
the Kavli Foundation.
The authors gratefully acknowledge the support of the NSF, STFC, MPS, INFN, CNRS and the
State of Niedersachsen/Germany for provision of computational resources.

This  article  has  been  assigned the document number P1600129.

\bibliographystyle{unsrt}
\bibliography{O1_allsky}

\end{document}